\documentclass[sigconf,screen]{acmart}

\setcopyright{cc}
\setcctype{by}
\acmDOI{10.1145/3832783.3837536}
\acmYear{2026}
\copyrightyear{2026}
\acmISBN{979-8-4007-2882-2/2026/10}
\acmConference[ASE '26]{Proceedings of the 41st IEEE/ACM International Conference on Automated Software Engineering}{October 12--16, 2026}{Munich, Germany}
\acmBooktitle{Proceedings of the 41st IEEE/ACM International Conference on Automated Software Engineering (ASE '26), October 12--16, 2026, Munich, Germany}
\acmSubmissionID{ase26main-p2547-p}
\received{2026-03-26}
\received[accepted]{2026-06-18}



\usepackage[utf8]{inputenc}
\usepackage[T1]{fontenc}

\usepackage{amsmath,amssymb,amsfonts,amsthm}
\usepackage{array}
\usepackage{graphicx}
\usepackage{textcomp}
\usepackage{multirow}
\usepackage{colortbl}
\usepackage[most]{tcolorbox}
\usepackage{xspace}
\usepackage{listings}
\usepackage{adjustbox}
\usepackage{makecell}
\usepackage{pifont}
\usepackage{microtype}

\usepackage{wasysym}
\usepackage{soul}
\usepackage{enumitem}
\usepackage{enumerate}
\usepackage{hyperref}

\usepackage{algorithm}
\usepackage{algpseudocode}

\usepackage{xcolor}
\newcommand{\revised}[1]{#1}

\newcommand{\tool}{\textsc{PropGen}\xspace}

\newcommand{\etal}{\hbox{\emph{et al.}}\xspace}
\newcommand{\eg}{\hbox{\emph{e.g.}}\xspace}
\newcommand{\ie}{\hbox{\emph{i.e.}}\xspace}

\newcommand{\wrt}{\hbox{\emph{w.r.t.}}\xspace}

\newcommand{\findbugs}{25\xspace}

\newcommand{\omninotes}{\emph{OmniNotes}\xspace}
\newcommand{\amaze}{\emph{Amaze}\xspace}
\newcommand{\files}{\emph{MaterialFiles}\xspace}
\newcommand{\markor}{\emph{Markor}\xspace}
\newcommand{\uhabits}{\emph{uhabits}\xspace}
\newcommand{\myexpense}{\emph{MyExpenses}\xspace}
\newcommand{\newpipe}{\emph{NewPipe}\xspace}
\newcommand{\retromusic}{\emph{RetroMusic}\xspace}
\newcommand{\outertune}{\emph{OuterTune}\xspace}
\newcommand{\orgzly}{\emph{Orgzly}\xspace}

\newcommand{\anki}{\emph{AnkiDroid}\xspace}
\newcommand{\antennapod}{\emph{AntennaPod}\xspace}

\newcommand{\kea}{\textsc{Kea}\xspace}
\newcommand{\app}{$\mathcal A$\xspace}

\begin{document}

\title{PropGen: Automated Property Generation for Property-Based Testing of Mobile Apps}

\author{Yiheng Xiong}
\orcid{0009-0006-9467-4891}
\affiliation{%
  \institution{Singapore Management University}
  \city{Singapore}
  \country{Singapore}
}
\email{yihengx98@gmail.com}

\author{Shiwen Song}
\authornote{Corresponding author.}
\orcid{0009-0008-7885-1135}
\affiliation{%
  \institution{Singapore Management University}
  \city{Singapore}
  \country{Singapore}
}
\email{swsong@smu.edu.sg}

\author{Bo Ma}
\orcid{0009-0003-5930-1246}
\affiliation{%
  \institution{East China Normal University}
  \city{Shanghai}
  \country{China}
}
\email{boma@stu.ecnu.edu.cn}

\author{Ting Su}
\orcid{0000-0003-1628-9796}
\affiliation{%
  \institution{East China Normal University}
  \city{Shanghai}
  \country{China}
}
\email{tsu@sei.ecnu.edu.cn}

\author{Xiaofei Xie}
\orcid{0000-0002-1288-6502}
\affiliation{%
  \institution{Singapore Management University}
  \city{Singapore}
  \country{Singapore}
}
\email{xfxie@smu.edu.sg}

\begin{abstract}
Mobile apps often suffer from functional bugs that do not cause crashes but instead manifest as incorrect behaviors under specific user interactions. Such bugs are difficult to detect by conventional automatic testing techniques because they often lack explicit \textit{test oracles}. Property-based testing can effectively expose them by specifying intended behavior as properties and checking them under diverse interactions. However, its practical use is limited by the reliance on manually written properties, which are difficult and expensive to construct.

To address this limitation, this paper explores the use of large language models (LLMs) to automate property construction for property-based testing of mobile apps. This is challenging in two ways. \textit{First}, it is difficult to systematically uncover and execute diverse app functionalities. \textit{Second}, it is difficult to derive valid properties from functionality execution results.
To address these challenges, we introduce \tool, which infers candidate app functionalities as hypotheses from GUI states, validates each hypothesis by executing it to collect behavioral evidence, synthesizes properties from the collected evidence, and refines imprecise properties based on testing feedback.
We implemented \tool and evaluated it on 12 real-world Android apps.
The results show that \tool can effectively identify and execute app functionalities, generate valid properties, and refine most imprecise ones.
Across all apps, \tool inferred 1,210 valid functionalities and correctly executed 977 of them, compared with 491 and 187 for the baseline.
It generated 985 properties, 912 of which were valid, and successfully refined 118 of 127 imprecise ones exposed during testing.
Using the resulting properties, we found \findbugs previously unknown functional bugs, many of which were missed by existing testing techniques.
\end{abstract}

\maketitle

\section{Introduction}
Mobile apps are highly interactive and stateful systems whose functionalities are largely driven by user interface interactions. Despite extensive testing efforts, functional bugs (\eg, incorrect interaction logic) remain prevalent in real-world apps~\cite{xiong2023empirical}. These bugs often do not manifest as crashes, making them difficult to detect using conventional GUI testing techniques that mainly emphasize code coverage or crash discovery~\cite{su_stoat_2017,dong2020time,gu2019practical,wang2020combodroid,machiry2013dynodroid,mao2016sapienz,wang2025llmdroid,liu2024make}. Manual testing (\eg, writing GUI tests) is widely used in practice to validate the functional correctness of mobile apps~\cite{linares2017developers,kochhar2015understanding}. However, it is brittle, costly to maintain, and typically covers only pre-defined \textit{happy paths}, often missing non-trivial functional bugs~\cite{kea}.

Property-based testing (PBT) offers a promising direction for addressing this challenge~\cite{claessen2000quickcheck}. Recent work has demonstrated that carefully designed properties can reveal non-trivial functional bugs that are missed by other techniques~\cite{kea,sun2024property,liang2026kea2}. In mobile app testing, developers specify expected behaviors as properties, and PBT automatically generates a large number of GUI events to explore diverse GUI states and check whether these properties hold. Compared with validating only a fixed set of manually crafted GUI test cases, this paradigm provides a more efficient way to assess functional correctness across diverse GUI states. 

However, the effectiveness of PBT fundamentally depends on the availability of high-quality properties, whose manual construction remains a major barrier to practical adoption~\cite{goldstein2024property,goldstein2022some}. This challenge is especially pronounced for mobile apps, where explicit behavioral specifications are often unavailable. As a result, testers must manually understand app functionalities, abstract their expected behaviors into executable properties, and refine the properties when reported violations are false positives. This manual and iterative process substantially limits the broader adoption of PBT.

To address this limitation, a promising direction is to leverage the reasoning and code-generation capabilities of Large Language Models (LLMs) to automate property construction. However, directly asking an LLM to generate properties is often unreliable: mobile apps lack explicit behavioral specifications, and without concrete evidence of how an app actually behaves, LLMs tend to hallucinate properties that the app never satisfies. Instead, our idea is to anchor property construction in the app's own functionalities: we collect the functionalities of the app, and generate the properties they are expected to satisfy.

\noindent{\textbf{Challenges.}} Realizing this idea faces two key challenges.
\textit{First}, how to systematically discover the functionalities of an app? Mobile apps rarely come with specifications: what an app can do, and how, is implicit in its user interface rather than stated anywhere. Its functionalities are further scattered across many GUI states, and among the enormous number of possible event sequences, only a few complete a functionality. Those that remain undiscovered yield no properties, leaving the corresponding app behavior unchecked.
\textit{Second}, how to generate properties from the discovered functionalities? A functionality reveals only what the app can do, whereas a property asserts what must hold every time it is used. Generating properties thus requires closing the gap between observed and expected behavior, and no specification is available to tell whether it has been closed correctly.

\noindent{\textbf{Our approach.}} To address the first challenge, we design a functionality hypothesis-guided exploration strategy that systematically uncovers app functionalities and executes them to collect behavioral evidence, \ie, concrete records of how each functionality is carried out. Given a GUI state, it infers candidate functionalities as hypotheses, grounding each of them in a concrete actionable widget so that the hypothesis is directly executable.
Each hypothesis is then validated by execution: the app is driven to carry out the hypothesized functionality, and the resulting trace both confirms the hypothesis and provides the behavioral evidence.
During this process, it incrementally expands the functionality pool when new functionalities are inferred, reuses previously inferred functionalities across recurrent GUI states, and falls back to lightweight random exploration only when no unexecuted functionality remains in the current GUI state. 

To address the second challenge, instead of generating properties directly from execution results, our approach first abstracts each functionality execution trace into a compact condition--event--outcome representation, derives a structured property description in natural language, and then translates it into an executable property. 
Because properties inferred from observed behavior may still be imprecise, it refines those that trigger false positives during testing. By grounding each refinement in the behavioral evidence from which the property was derived, it localizes the imprecision to one of three components, the \textit{precondition}, \textit{interaction}, or \textit{postcondition}, and revises only that component. This makes the generated properties more precise while preserving their original intent. 

\noindent{\textbf{Evaluation and results.}} We implemented our approach as a tool named \tool and evaluated it on 12 real-world popular and diverse Android apps. 
Across all apps, \tool inferred 1,282 functionalities, of which 1,210 (94.4\%) were valid and 977 (76.2\%) were correctly executed. In comparison, the baseline approach inferred 575 functionalities, with 491 valid and 187 correctly executed.
It further generated 985 property descriptions, 912 (92.6\%) of which were valid. 
During property-based testing, 127 properties were found to be imprecise, and 118 (92.9\%) of them were successfully refined by our refinement technique. 
Using the generated properties, we found \findbugs previously unknown functional bugs in the latest versions of the subject apps, whereas existing functional testing techniques could find only 3 of them in practice. 
These results demonstrate the effectiveness of our approach in automating property construction for mobile apps, as well as the bug-finding capability of the resulting properties.

In summary, this paper makes the following contributions:
\begin{itemize}[leftmargin=*]

    \item At the conceptual level, we propose a novel approach that automatically explores app functionalities and generates properties from runtime behavioral evidence, without relying on manual specifications.

    \item At the technical level, we design a hypothesis-guided exploration technique that infers functionalities from GUI states and collects behavioral evidence from their executions, together with a property synthesis technique that derives properties from such evidence and refines imprecise ones through testing feedback.

    \item At the empirical level, we implement our approach as \tool and evaluate it on 12 real-world Android apps. \tool generates 912 valid properties, refines 92.9\% of the imprecise ones, and uncovers \findbugs previously unknown functional bugs.

\end{itemize}

\begin{figure*}[t]
  \centering
  \includegraphics[width=0.975\linewidth]{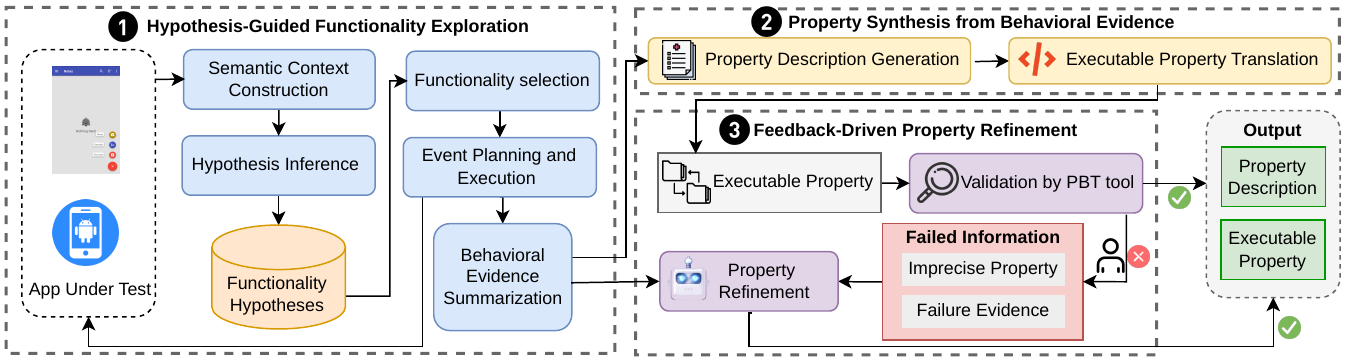}
  \caption{Overview of \tool.}
  \label{fig:overview}
\end{figure*}
\begin{figure*}[t]
  \centering
  \includegraphics[width=0.975\linewidth]{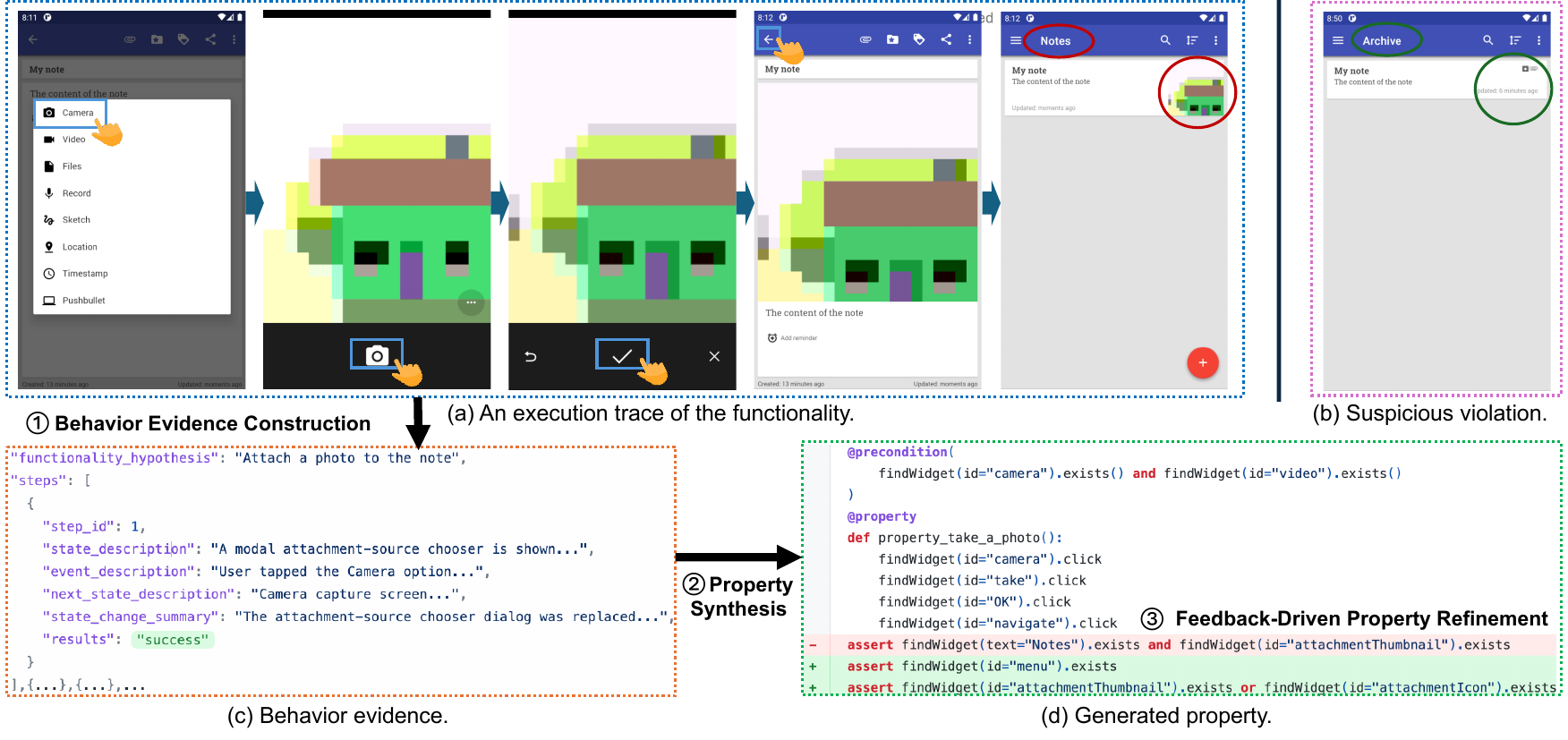}
  \caption{An Illustrative Example of Behavioral Evidence Construction, Property Synthesis, and Feedback-Driven Property Refinement for a Note-Taking App.}
  \label{fig:example}
\end{figure*}

\vspace{-10pt}
\section{Background}

\textbf{Property-based testing.}
Property-based testing is a powerful testing methodology that validates whether a program satisfies general properties rather than specific input-output examples~\cite{claessen2000quickcheck}. Instead of writing example-based test cases, testers specify high-level properties that describe the expected behavior of the system. Then, a PBT framework automatically generates a large number of test inputs and executes them to verify whether the properties hold.
For example, for a sorting function \texttt{sort}, rather than enumerating concrete examples (\eg, \texttt{sort([3,1,2])}$==$ \texttt{[1,2,3]}), one can define a general property such as idempotence: \texttt{sort(sort(x)) = sort(x)}. The PBT tool then generates a large number of inputs to verify this property and reports any violation as a counterexample.

\label{sec:problem-definition}

\noindent{\textbf{GUI state, event, and functionality.}}
Android applications are GUI-driven and event-based. When an app \app runs, its runtime state is represented by its current GUI layout, which we denote as a \emph{GUI state} \(s\). A GUI state corresponds to a hierarchical tree \(\ell\), whose nodes are GUI widgets \(w\) (\eg, \texttt{Button}, \texttt{TextView}, \texttt{EditText}) with attributes (\eg, \texttt{text}, \texttt{resourceId}) and interaction capabilities.
User interactions are modeled as \emph{events}. An event is defined as \(e = \langle t, w, v \rangle\), where \(t\) is the event type (\eg, click), \(w\) is the target widget, and \(v\) is optional data (\eg, text input).
An app execution is modeled as a sequence of events. Given \(E = [e_1, \ldots, e_n]\), executing \app produces a trace \(\tau = s_0 \xrightarrow{e_1} s_1 \xrightarrow{} \cdots \xrightarrow{e_n} s_n\), or \(s_0 \overset{E}{\rightsquigarrow} s_n\), where \(s_0\) is the initial state.

\revised{We define a \emph{functionality} as a tuple $f=\langle d,T \rangle$, where $d$ is a natural-language description of the semantic intent of the functionality (\eg, ``create a note''), and $T=\{\tau_1,\tau_2,\ldots\}$ is the set of execution traces that realize this intent. 
Executing any $\tau \in T$ completes the functionality and changes the app state.
These traces share the intent $d$ but may differ in input data, events, and starting GUI states.
We refer to what can be observed when a trace $\tau \in T$ is executed, \ie, the GUI states it traverses and the changes it induces, as a \emph{behavior} of $f$. Each trace thus witnesses one behavior of the same functionality, and such observed behaviors form the \emph{behavioral evidence} from which properties are derived in this paper.}

\noindent{\textbf{Property-based testing for mobile apps.}}
A property is defined as a tuple
$
\phi = \langle P, I, Q \rangle,
$
where the \emph{precondition} $P$ characterizes the GUI states in which the property applies, $I$ is the \emph{interaction scenario}, \revised{\ie, a sequence of statements that defines how to interact with the app, which may also contain control-flow constructs such as conditionals and loops,} and the \emph{postcondition} $Q$ specifies the expected outcome after executing $I$.
During testing, executing $I$ from a GUI state $s$ that satisfies $P$ leads to a new state $s'$, and the property holds if
$
\forall s.\ (s \models P \wedge s \overset{I}{\rightsquigarrow} s') \Rightarrow s' \models Q.
$
Otherwise, a property violation is reported.

\section{Approach}
\noindent{\textbf{Overview.}}
Given an app, \tool automatically generates properties from runtime executions and further refines imprecise ones based on testing feedback. Figure~\ref{fig:overview} presents the overall workflow, which consists of three stages. First (\S\ref{sec:evidence}), \tool performs hypothesis-guided dynamic exploration to construct behavioral evidence. For each encountered GUI state $s$, it infers functionality hypotheses grounded in the visible UI widgets, executes selected functionalities, and summarizes the resulting execution traces into behavioral evidence for downstream property generation. 
Second (\S\ref{sec:property-synthesis}), \tool synthesizes properties directly from the behavioral evidence constructed in the first stage. Specifically, it first generates natural-language property descriptions that capture the intended condition--event--outcome relation, and then translates them into executable properties.
Third (\S\ref{sec:refinement}), \tool validates the generated properties with the PBT framework, and refines those whose reported violations are manually confirmed to be false positives rather than real bugs.

\noindent{\textbf{Illustrative example.}}
Figure~\ref{fig:example} illustrates the workflow of \tool on a note-taking app. From the page in Figure~\ref{fig:example}(a), \tool infers candidate functionality hypotheses and selects one of them, \ie, attaching a photo to the note. It then executes the selected hypothesis by opening the attachment menu, choosing Camera, taking a photo, and returning to the note page. The resulting trace is summarized into structured behavioral evidence (Figure~\ref{fig:example}(c)), from which \tool synthesizes an executable property for this functionality (Figure~\ref{fig:example}(d)).

This property, however, is imprecise. Testing it with the PBT framework \kea reports violations that manual inspection confirms to be false positives: the postcondition requires the app to return to a page containing the text “Notes” and the attachment to appear as a thumbnail, but the app may legitimately return to either the Notes or the Archive page, depending on where the note was opened, and under reduced view the attachment appears as a compact icon (Figure~\ref{fig:example}(b)). Guided by this feedback, \tool relaxes both assertions to accept either return page, checked through a menu button that both contain, and either attachment representation.
Re-testing then uncovers a new functional bug: opening audio recording before taking a photo prevents the photo attachment from appearing.

\subsection{Hypothesis-Guided Functionality Exploration}
\label{sec:evidence}
The goal of this stage is to construct execution-grounded behavioral evidence for downstream property synthesis. 
Instead of relying on external specifications, \tool systematically explores executable app functionalities and records their runtime interaction traces as behavioral evidence. 
Algorithm~\ref{alg:evidence} summarizes the workflow.

Given a target app $\mathcal{A}$ and a time budget $B$, \tool launches the app, initializes the global functionality pool $\mathcal{G}_F$, the set of observed UI contexts $\mathcal{U}$, and the behavioral evidence set $\hat{\mathcal{T}}$ (Lines~1--2), and then iteratively explores the app until the budget is exhausted (Lines~3--19).
In each iteration, it extracts the current GUI context (Line~4); if the context contains previously unseen widgets, it invokes a Multimodal Large Language Model (MLLM) to infer functionality hypotheses for the current state and adds them to $\mathcal{G}_F$ (Lines~5--8).
It then executes an unexplored hypothesis together with its triggering widget, and summarizes the resulting trace into a behavioral evidence item added to $\hat{\mathcal{T}}$ (Lines~10--15).
When no unexplored hypothesis remains in the current context, \tool performs lightweight random exploration to leave the current local GUI region and expose new interaction opportunities (Lines~16--17).

\begin{algorithm}[t]
\caption{Behavioral Evidence Construction}
\label{alg:evidence}
\begin{algorithmic}[1]
\Require Target app \(\mathcal{A}\), time budget \(B\)
\Ensure Behavioral evidence set \(\hat{\mathcal{T}}\)

\State Launch \(\mathcal{A}\); obtain initial state \(s\)
\State \(\mathcal{G}_F \gets \emptyset\), \(\mathcal{U} \gets \emptyset\), \(\hat{\mathcal{T}} \gets \emptyset\)

\While{elapsed time $< B$}
    \State \(u(s) \gets \textsc{Extract}(s)\) \Comment{Extract the widgets}

    \If{\textsc{UnseenWidgets}\((u(s), \mathcal{U})\)}
        \State \(F(s) \gets \textsc{Inferhypothesis}(s)\) \Comment{Infer the hypotheses}
        \State \(\mathcal{G}_F \gets \mathcal{G}_F \cup F(s)\) \Comment{Add hypotheses to global pool}
        \State  \(\mathcal{U}\ \gets \mathcal{U} \cup u(s) \)
    \EndIf

    \If{\textsc{HasUnexplored}\((\mathcal{G}_F(u(s)))\)}
        \State \(f \gets \textsc{Select}(\mathcal{G}_F(u(s)))\)
        \State \((\tau_f, s', \mathcal{G}_F, \mathcal{U}) \gets \textsc{Execute}(f,s,\mathcal{G}_F,\mathcal{U})\)
        \State \(\hat{\mathcal{T}} \gets \hat{\mathcal{T}} \cup \{\textsc{Summarize}(\tau_f)\}\) \Comment{Summarize evidence}
        \State \(\textsc{MarkExplored}(f,\mathcal{G}_F)\)
        \State \(s \gets s'\)
    \Else
        \State \(s \gets \textsc{RandomExplore}(s)\) 
    \EndIf
\EndWhile

\State \Return \(\hat{\mathcal{T}}\)

\end{algorithmic}
\end{algorithm}

\subsubsection{Functionality Hypothesis Generation}
\label{sec:hypothesis_generation}

Given a GUI state \(s\) encountered during exploration, this step aims to infer which app functionalities may be executable under the current interface context. To support subsequent execution, each inferred functionality is associated with a concrete \emph{triggering widget} in the current state, \revised{\ie, the widget from which the functionality is initiated (\eg, a ``+'' button initiates ``create a note'').}

\noindent{\textbf{Semantic context construction.}} 
To support functionality inference, \tool first constructs a semantic context for the current GUI state \(s\). This context includes three parts: the screen-level context of the current state, app-level semantic cues, and cross-state functionality memory. 
The screen-level context is derived from the current GUI screenshot. \revised{To obtain explicit widget references, \tool retrieves the accessibility tree of \(s\) via uiautomator2~\cite{uiautomator2}, keeps the leaf-level interactive widgets, and draws a sequential number at each widget's top-right corner using its \texttt{bounds} attribute. These numbers let the MLLM name a widget rather than describe it in prose, grounding the inferred functionalities in concrete widgets. They are recomputed per screen and carry no meaning across screens; cross-screen identity is instead established by the widget signature described below.}
The app-level semantic cues include the app name and the list of activity names extracted from the \texttt{AndroidManifest.xml} file, helping the MLLM interpret the current screen under the broader semantic context of \app. 
Finally, the cross-state functionality memory stores previously inferred functionalities, helping avoid repeatedly rediscovering semantically similar functionalities in later GUI states. Together, these components provide the contextual information needed for inferring plausible user-facing functionalities from \(s\).

\noindent{\textbf{Functionality hypothesis inference.}}
Using the constructed semantic context, \tool invokes an MLLM to infer candidate functionality hypotheses for the current GUI state \(s\). Formally, the inferred hypotheses are represented as
$
F(s) = [\langle f_1, w_1\rangle, \langle f_2, w_2\rangle, \ldots,$ $\langle f_k, w_k\rangle],
$
where each functionality hypothesis \(f_i\) is a concise natural-language description of a candidate app functionality, and \(w_i\) denotes its corresponding triggering widget. This widget grounding serves two purposes. First, it constrains the MLLM to infer functionalities that are supported by the current GUI state, reducing unsupported or non-actionable predictions. Second, it makes the inferred functionalities directly executable in subsequent exploration. The inferred functionality hypotheses, together with their associated triggering widgets, are then stored in the global functionality pool \(\mathcal{G}_F\) to support later execution and cross-state reuse.

\noindent{\textbf{Inference triggering and hypothesis reuse.}}
To avoid redundant MLLM invocations on similar GUI states, \tool triggers functionality inference only when a newly visited state introduces unseen widgets. It maintains a global set of explored UI contexts
\(
\mathcal{U} = \{u_1, u_2, \ldots\},
\)
where each context is defined as
\(
u(s) = \langle a(s), W(s) \rangle,
\)
with \(a(s)\) denoting the activity and \(W(s)\) the set of signatures of leaf-level interactive widgets.

For each state \(s\), \tool compares \(W(s)\) with existing contexts under the same activity. If no new widget signatures are observed, it reuses previously inferred functionality hypotheses; otherwise, it invokes the MLLM to infer new functionalities and updates the global pool \(\mathcal{G}_F\).
To enable stable comparison across screens, each widget \(w\) is represented by a signature of attributes
$
 \langle class, resourceId, text, description \rangle,
$
which are commonly used to identify unique widgets in practice~\cite{su_stoat_2017,pbfdroid,wen2023droidbot}. To reduce noise from dynamic content, \tool retains only app-defined text in signatures by filtering widget text against a whitelist extracted via static analysis. This avoids treating semantically identical screens with transient text differences as distinct contexts.

\subsubsection{Functionality Hypothesis Validation}

Given the global functionality pool \(\mathcal{G}_F\) generated in \S~\ref{sec:hypothesis_generation}, \tool executes app functionalities in a targeted manner rather than through arbitrary GUI events: it treats each hypothesis \(\langle f, w\rangle\) in \(\mathcal{G}_F\) as an explicit execution target and prioritizes previously unexplored ones. Executing \(\langle f, w\rangle\) induces state transitions \(s \xrightarrow{E} s'\), which both collect behavioral evidence for the executed functionality and drive the app into new states \(s'\), where further hypotheses \(F(s')\) are inferred and added to \(\mathcal{G}_F\). Execution and hypothesis generation thus form a closed exploration loop.

\noindent{\textbf{Functionality selection.}} 
At each GUI state \(s\), \tool retrieves from \(\mathcal{G}_F\) the candidate functionality hypotheses associated with the current UI context \(u(s)\), denoted as \(\mathcal{G}_F(u(s))\). Since the exploration budget is limited, \tool does not execute these candidates arbitrarily, but prioritizes those with higher expected exploration utility. This prioritization is implemented using lightweight heuristics guided by three considerations: whether the functionality corresponds to a main app behavior, whether it is semantically different from previously executed functionalities, and whether it is likely to be successfully executed in the current context. \revised{These considerations are operationalized by prompting the MLLM to rank the candidates accordingly.} Based on these heuristics, the top-ranked unexplored functionality hypothesis \(\langle f, w\rangle\) is selected as the next execution target. The selected hypothesis is then passed to a goal-directed interaction loop that incrementally plans, executes, and evaluates UI actions until the functionality is completed or a step limit is reached.

\noindent{\textbf{Event planning and execution.}} 
Once a functionality hypothesis \(\langle f, w\rangle\) is selected, \tool executes it through a goal-directed interaction loop consisting of event planning, guarded execution, and post-event evaluation. At each step \(i\), \tool first predicts the next GUI event based on three sources of information: the functionality goal \(f\), the execution history \(H_i\) of prior events and their outcomes, and the current GUI state \(s_i\), represented as a screenshot with labeled widgets. The predicted event  is denoted as \(e_i = \langle t_i, w_i, v_i \rangle\), where the event  type \(t_i\) is chosen from a predefined event  space including \texttt{click}, \texttt{long-click}, \texttt{edit}, \texttt{swipe}, and \texttt{back}, the target widget \(w_i\) is specified by its numeric label, and \(v_i\) specifies the attached data.
Before executing \(e_i\), \tool verifies that the referenced widget \(w_i\) is present in the current state \(s_i\). If the predicted widget identifier is invalid, the event is skipped and directly labeled as a failed step. This guarded execution mechanism prevents invalid model-predicted interactions and improves execution robustness.

After execution produces a state transition \(s_i \xrightarrow{e_i} s_{i+1}\), \tool evaluates whether the step has advanced the selected functionality. This evaluation jointly considers the pre- and post-event GUI states, the functionality hypothesis \(f\), and the accumulated execution history \(H_i\). Based on this evidence, the MLLM assigns an outcome label
$
o_i \in \{\textit{success}, \textit{fail}, \textit{complete}\}
$
to the current step. Here, \textit{success} indicates meaningful progress toward the functionality goal, \textit{fail} indicates that the event was unproductive for the intended functionality, and \textit{complete} indicates that the functionality goal has been achieved. The resulting step and outcome are then incorporated into the execution history for subsequent planning, while \textit{complete} terminates execution of the current functionality.

\noindent{\textbf{Behavioral evidence summarization.}} 
To support downstream property synthesis, \tool summarizes each executed interaction trace into compact behavioral evidence, since raw traces are low-level, noisy, and specific to a single execution. It incrementally converts each interaction step \(s_i \xrightarrow{e_i} s_{i+1}\) into a structured transition with five elements: the state summary before the event, an event summary describing the interaction performed, the state summary after the event, a state-diff summary capturing the visible difference between the two GUI states, and the outcome label \(o_i\).

The state summaries are functionality-oriented: they describe the screen context, visible actionable elements, current content state, and observable feedback cues, whereas the state-diff summary focuses on the concrete GUI changes induced by the interaction. The resulting summarized trace, denoted as \(\hat{\tau}_f\), serves as the behavioral evidence used in the subsequent property synthesis stage.

\noindent{\textbf{Exploration beyond local hypothesis exhaustion.}}
\tool may reach a GUI state whose functionality hypotheses have all been executed. The only remaining task is to move the app into a new GUI region, for which querying the MLLM would cost tokens and time without yielding new behavioral evidence. \tool therefore falls back to lightweight random exploration, issuing random events until it reaches a state that either has an unexecuted hypothesis, which it executes directly, or unseen widgets, for which it first infers new hypotheses.

\vspace{-5pt}
\subsection{Property Synthesis from Behavioral Evidence}
\label{sec:property-synthesis}

The goal of this stage is to derive executable properties from the behavioral evidence collected during functionality exploration. As input, \tool takes the summarized trace \(\hat{\tau}_f\) produced in Section~\ref{sec:evidence}, which compactly captures how a functionality is exercised and what observable GUI outcome it induces. Each synthesized property takes the form
$
\phi = \langle P, I, Q \rangle,
$
where \(P\) is a precondition, \(I\) is an interaction scenario, and \(Q\) is a postcondition assertion.

Rather than generating \(\phi\) directly from \(\hat{\tau}_f\), \tool decomposes property synthesis into two steps, separating semantic property formulation from executable code generation. It first constructs a natural-language property specification that explicitly captures the intended \(\langle P, I, Q \rangle\) relation, and then translates this intermediate specification into executable property code.

\noindent{\textbf{Natural-language property description generation.}}
Based on the summarized trace \(\hat{\tau}_f\), \tool first prompts the MLLM to construct a natural-language property specification
$
\phi^{NL} = \langle P, I, Q \rangle
$
for the functionality captured by the behavioral evidence. The summarized trace provides structured evidence about the execution context, performed interactions, observable state changes, and step outcomes. Rather than merely restating these observations, the MLLM is guided to abstract from \(\hat{\tau}_f\) a generalized behavioral rule that should hold across executions of the same functionality.

The inferred specification \(\phi^{NL}\) consists of three parts: a \emph{precondition} \(P\) describing the observable GUI context in which the property should be checked, an \emph{interaction scenario} \(I\) capturing the user events needed to exercise the functionality, and a \emph{postcondition} \(Q\) specifying the immediate visible effect that should hold afterward. To improve precision and executability, \tool constrains this inference process in three ways. First, \(P\) must be grounded in observable UI evidence and sufficiently specific to avoid triggering the property on unrelated screens with superficially similar widgets. Second, \(Q\) must focus on effects that are directly and reliably verifiable from the GUI, such as the appearance, disappearance, or modification of visible widgets or content. Third, the inferred specification should avoid trace-specific brittle details, such as incidental text instances or unstable widget states, and instead capture functionality semantics in a form that remains robust across executions.
\revised{These three constraints are enforced as explicit instructions in the synthesis prompt.}

\noindent{\textbf{Executable property translation.}}
Given the inferred natural-language property specification
$
\phi^{NL} = \langle P, I, Q \rangle,
$
\tool translates it into executable property code for the target PBT framework. Specifically, this step realizes the precondition \(P\), interaction scenario \(I\), and postcondition \(Q\) using the framework's property structure and API conventions, thereby producing a runnable property implementation. Since the intended property semantics have already been made explicit in the preceding step, this translation primarily serves to operationalize the structured specification rather than to perform further behavioral inference.

This step is implemented by prompting the LLM with the generated natural-language specification together with framework-specific APIs and widget attributes, with reference to the prompt design in prior work on translating natural-language properties into executable ones~\cite{xiong2026naturallanguageexecutableproperties}. The resulting code is then passed to the subsequent validation and refinement stage.

\vspace{-5pt}
\subsection{Feedback-Driven Property Refinement}
\label{sec:refinement}

Properties automatically generated from functionality traces may still be imprecise and therefore trigger false positives during testing.
However, a reported failure should not be refined by simply adapting the property to that single execution, because such a fix may overfit the observed case and drift away from the original functionality intent.

Given a property whose reported violations are manually confirmed to be false positives, our key idea is that refining it requires first recovering the property's original testing intent, and then revising it so that it remains valid for both the original intended execution and the newly observed legitimate execution. 
To this end, we ground refinement in the source summarized trace from which the property was originally inferred, rather than treating refinement as unconstrained rewriting.

Specifically, \tool reasons about the refinement using both the source summarized trace and the failure-triggering execution. 
It first recovers the original testing intent of the property by locating the relevant segment in $\hat{\tau}_f$ that matches the property's triggering context, interaction scenario, and expected outcome. 
It then compares this trace-grounded intended behavior with the failing execution to explain why the violation occurs and identify which component of $\phi$ has become imprecise, \ie, the precondition $P$, the interaction scenario $I$, or the postcondition $Q$.

Based on this diagnosis, \tool refines only the faulty component through a minimal modification. 
More specifically, it strengthens $P$ when additional UI guards are needed, revises $I$ when the event sequence does not faithfully reflect the intended behavior, and relaxes or simplifies $Q$ when the assertion is overly specific. 
In this way, the refined property remains consistent with both the source-trace execution and the newly observed legitimate execution, while preserving the original testing intent as much as possible.


\section{Implementation}
We implemented \tool as an end-to-end prototype for automated property generation on Android apps.
\tool is built on top of DroidBot~\cite{li2017droidbot}, an open-source Android app exploration tool. It is primarily written in Python and JavaScript. Specifically, \tool uses uiautomator2~\cite{uiautomator2} to retrieve GUI layouts and Android Debug Bridge (ADB)~\cite{adb} to capture screenshots and send GUI events, including \texttt{click}, \texttt{long-click}, \texttt{edit}, \texttt{swipe}, and \texttt{back}. The generated properties are translated into the executable format expected by \kea. 

\section{Evaluation}

We aim to answer the following research questions:

\begin{itemize}[label=\textbullet,leftmargin=*]
    \item \textbf{RQ1}: How effective is \tool in exploring app functionalities and generating valid properties?

    \item \textbf{RQ2}: To what extent do generated properties suffer from imprecision that leads to false positives, and how effective is our refinement technique in refining them?
    
    \item \textbf{RQ3}: Can the generated properties help find new functional bugs in real-world mobile apps?

    \item \textbf{RQ4}: How does \tool compare with prior functional testing techniques in uncovering new functional bugs?
\end{itemize}

\begin{table}[!t]
    \centering
    \small
    \caption{App subjects used in our experiment (K=1,000, M=1,000,000).}
    \label{tab:alltheapps}
    \renewcommand{\arraystretch}{1.2} 
    \setlength{\tabcolsep}{11pt} 
    
    \begin{adjustbox}{max width=0.48\textwidth} 
        \begin{tabular}{l|l|r|r|r}
            \hline
            \hline
            \rowcolor[HTML]{EFEFEF} 
            \textbf{App Name} & \textbf{App Feature} & \textbf{\#Downloads} & \textbf{\#Stars} & \textbf{LOC} \\ 
            \hline

            \omninotes    & Note Manager        & 100$\sim$500K       & 2.8K    & 57,529    \\
            \markor       & Text Editor         & -                   & 5.3K    & 79,749    \\

            \retromusic   & Audio Player        & 1$\sim$5M           & 5K      & 110,065   \\
            \amaze        & File Manager        & 1$\sim$5M           & 6.1K    & 159,040   \\

            \myexpense    & Financial Assistant & 1$\sim$5M           & 1.1K    & 317,899   \\
            \antennapod   & Podcast Manager     & 1$\sim$5M           & 7.7K    & 130,925   \\

            \anki         & Flashcards Manager  & 10$\sim$50M         & 10.9K   & 403,785   \\
            \outertune    & YouTube Music Player & -                  & 4.8K    & 89,673    \\

            \newpipe      & Video Player        & -                   & 37.5K   & 187,187   \\
            \files        & Storage Browser     & 1$\sim$5M           & 8K      & 95,705    \\

            \orgzly       & To-do Lists Manager & 100$\sim$500K       & 2.8K    & 72,042    \\
            \uhabits      & Habit Tracker       & 5$\sim$10M          & 9.7K    & 69,652    \\
            
            \hline
            \hline
        \end{tabular}
    \end{adjustbox}
\end{table}

\subsection{Setup and Method}

\textbf{App subjects.} 
We selected 12 popular and representative open-source Android apps as experimental subjects. 
Among them, eight apps were adopted from prior studies on functional bug detection for Android apps~\cite{kea,genie,odin,sun2024property}. 
From the apps used in these studies, we excluded those that were either (1) no longer runnable or actively maintained, or (2) highly similar in functionality to apps already selected. 
To further improve diversity, we additionally included four popular open-source apps from Google Play with different features. 
Table~\ref{tab:alltheapps} summarizes the selected apps. In the table, \textit{App Feature} denotes the primary functionality of each app, \textit{\#Downloads} and \textit{\#Stars} report the number of Google Play installations and GitHub stars, respectively, and \textit{LOC} gives the lines of code.

\noindent\textbf{Experimental environment.}
All experiments were conducted on a machine running Ubuntu 22.04 with 192 CPU cores (AMD EPYC 9654) and official Android emulators (Android 11, Pixel). 
\revised{We use GPT-5.2 (snapshot \texttt{gpt-5.2-2025-12-11}) as the backend MLLM for \tool and the baseline tool, with the default decoding parameters (temperature $=1.0$, top-$p=1.0$).}
For each app, we allocated 3 hours for behavioral evidence construction and property synthesis. 
After that, we used \kea to perform property-based testing with the generated properties for 6 hours per app, following the testing budget adopted in \kea's paper~\cite{kea}.

\noindent\textbf{Baselines.}
We use six baselines for different research questions, according to evaluation goals. 
For RQ1, we compare \tool with DroidAgent~\cite{yoon2024intent}, a representative LLM-based mobile app functionality exploration approach, which can automatically identify and execute functionalities.
\revised{To assess whether the multi-stage pipeline of \tool is necessary at all, we additionally compare against a \emph{naive LLM} baseline that asks the LLM to generate properties directly, without any runtime exploration. Concretely, we prompt the same MLLM with the app's official description and ask it to produce 20 properties per app, using the same property format as \tool.}
For RQ4, which evaluates bug-finding capability, we compare \tool with four representative prior techniques for Android functional bug detection: Genie~\cite{genie}, Odin~\cite{odin}, PBFDroid~\cite{pbfdroid}, and VisionDroid~\cite{liu2025seeing}. 
Among them, Genie and Odin rely on designed automated oracles, PBFDroid is a property-based testing technique for data manipulation functionalities (DMFs), and VisionDroid is an LLM-based multi-agent approach for functional bug detection.

\noindent\textbf{Evaluation method of RQ1.}
RQ1 aims to evaluate whether \tool can correctly infer and execute app functionalities and synthesize valid properties from the collected behavioral evidence. 
This evaluation cannot be performed fully automatically, because mobile apps typically lack precise functional specifications. 
Therefore, we manually assess the validity of both the inferred functionalities and the generated property descriptions, following prior work on functionality-level evaluation beyond structural coverage metrics~\cite{coppola2022taxonomy,yoon2024intent}. 
The manual evaluation is conducted by two graduate students in software engineering, each with at least four years of Android development experience. \revised{Neither annotator is a co-author or was involved in developing \tool, which reduces assessment bias.}
Before the annotation, each annotator was given at least fifteen minutes to familiarize themselves with the functionalities of every subject app.
During this process, annotators also referred to the app's official introduction page when available, so that the subsequent judgments were made with sufficient understanding of the app's functionalities. In addition, annotators could interact with the running app at any time to verify uncertain cases.

For each inferred functionality, we evaluate two aspects. 
The first is \emph{functionality validity}, which examines whether the inferred functionality actually exists in the app. 
Annotators are given the inferred functionality description together with the GUI screenshot from which it was inferred, and determine whether the described functionality is genuinely supported by the interface. 
The second is \emph{execution correctness}, which examines whether the system correctly executes the inferred functionality. 
For each inferred functionality, annotators are provided with the corresponding execution screenshots and interaction events, and judge whether the executed interaction sequence indeed realizes the intended functionality. 

\revised{Because mobile apps lack ground-truth specifications, we cannot decide a property's validity against a reference oracle. We therefore assess validity along two complementary layers, one checked by an automated proxy and the other by human annotators.
The \emph{groundedness} layer is verified by an \emph{automated proxy}: we automatically check that every widget and event referenced in the property is observable in its source trace, so that the property is executable on the app rather than referring to hallucinated UI elements.
The \emph{faithfulness} layer is judged by \emph{human annotators}, who assess the three components:
(1) whether the \emph{precondition} appropriately constrains the UI state in which the property should be applied,
(2) whether the \emph{interaction scenario} accurately reflects the user interactions required to perform the functionality, and
(3) whether the \emph{postcondition} correctly captures the observable UI behavior that should hold immediately after the scenario.
A property description is considered \emph{valid} only if it is grounded in its source trace \emph{and} all three components are faithful to the intended functionality; otherwise, it is labeled as \emph{invalid}.
Validity does not imply precision: whether a property avoids false positives under arbitrary executions is a separate question, examined in RQ2.}

\revised{To reduce bias, the two annotators first labeled all results fully independently, without discussing their judgments with each other or with the authors. We measured inter-rater agreement on these independent annotations using Cohen's $\kappa$, obtaining 0.91, 0.82, and 0.81 for functionality validity, execution correctness, and property validity, respectively, which indicates high agreement across all three aspects.
Only after the independent annotation was complete did the two annotators discuss the remaining disagreements between themselves and reach a consensus label for each disagreed item.}
Based on these final annotations, we report the proportions of inferred functionalities that are valid, inferred functionalities that are correctly executed, and generated valid property descriptions.

\noindent\textbf{Evaluation method of RQ2 and RQ3.}
RQ2 evaluates two aspects of property refinement: (1) how many generated properties are imprecise and thus lead to spurious violations during testing, and (2) how many of these imprecise properties can be successfully refined through refinement.  RQ3 evaluates whether the generated properties can find new functional bugs.
Specifically, for each app, we load all initially generated executable properties into \kea~\cite{kea} for testing with 6 hours. 
Whenever \kea reports a property violation, we manually inspect the property description, executable property code, violation-triggering execution trace, corresponding screenshots and interaction events, and the assertion outcome. 
Based on this, we determine whether the violation is caused by a real app bug or by property imprecision.

For RQ2, we focus on the properties whose reported violations are diagnosed as spurious. 
We apply refinement only to those caused by property imprecision. 
For each such property, the refinement module takes the original property description, executable property code, and violation-triggering execution evidence as input, and produces a revised executable property. 
\revised{We then re-execute the refined property on two traces and inspect the results: the source trace from which the property was originally derived, and the trace that triggered the spurious violation.
We call the revised property \emph{plausibly correct} if it passes two automated checks: (a) it still holds on the source trace, and (b) it no longer reports the previously observed spurious violation on the violation-triggering trace.
These two checks form an \emph{automated proxy}: they are necessary but not sufficient, since a revision can pass both while still failing to faithfully specify the intended functionality.
We therefore reserve the term \emph{correct} for a plausibly correct property that human annotators further confirm to be \emph{valid} (\ie, grounded and faithful, as defined in RQ1) and to \emph{preserve the original testing intent}, by inspecting the full $\langle P,I,Q \rangle$ triple. A refinement is considered successful only when the revised property is both plausibly correct and human-confirmed correct.}
Based on this process, we report the number of properties sent to refinement due to property imprecision, the number and rate of successful refinements, and the breakdown of successfully refined properties by the modified component (\ie, precondition, interaction, and postcondition).

For RQ3, we focus on the violations diagnosed as real app bugs through manual inspection. 
For each confirmed bug, we prepare a bug report containing the bug description, reproduction steps, expected and actual behaviors, and submit it to the corresponding app developers.

\noindent\textbf{Evaluation method of RQ4.}
RQ4 compares \tool with existing functional testing techniques in uncovering new functional bugs.
\revised{Non-crash functional bug detection is \emph{oracle-dependent}: each technique can only expose the bugs that its oracle is able to express. Raw bug counts alone are therefore not directly comparable across these tools, as a tool that reports fewer bugs is not necessarily less capable, but may target a narrower class of bugs by design. To draw a meaningful comparison, we report absolute bug counts for completeness \emph{and} complement them with a scope analysis, which separates the two distinct reasons a tool can miss a bug: its oracle cannot express the bug at all, or its oracle can express the bug but the tool failed to expose it within the testing budget. Reporting both lets us measure overall bug-finding effectiveness while also explaining where the differences come from.}

\revised{Concretely, following prior comparative analysis practice~\cite{kea,pbfdroid}, we evaluate these tools from two perspectives.
First, we measure the \emph{overall bug-finding results} of each tool empirically:} we run each prior technique on the same 12 apps and count all previously unknown functional bugs it uncovers, \revised{using the same manual-confirmation criterion as for \tool, \ie, each reported violation is inspected and counted only if it is confirmed to be a real, previously unreported functional bug.}
For Genie, Odin, and VisionDroid, we follow the default configurations described in their original papers.
PBFDroid requires users to manually specify properties for data manipulation functionalities (DMFs); therefore, we manually defined the required DMF properties for detecting the corresponding bugs.
To ensure fairness, we align the time budget with each tool's workflow: \tool uses 3 hours for property generation and 6 hours for bug finding; accordingly, we allocate 9 hours per app to Genie, Odin, and VisionDroid, and 6 hours of automated testing to PBFDroid after manual property construction.
\revised{This single run yields both each tool's total bug count, independent of the bug set found by \tool, and, by intersecting with \tool's \findbugs new bugs, how many of them each tool rediscovers in practice.}

\revised{Second,} we conduct a scope analysis to determine whether each of the \findbugs bugs uncovered by \tool theoretically falls within the detection scope of each prior technique.
This analysis is performed manually based on bug characteristics and the detection capabilities claimed by each technique.
To improve reliability, we further consulted the authors of Genie, Odin, and PBFDroid to validate our analysis.
For VisionDroid, we do not perform a separate scope analysis, since its LLM-based design makes its theoretical detection scope difficult to characterize precisely; instead, we rely only on its empirical bug-finding performance.

\begin{table}[!t]
\centering
\caption{Validity of functionalities and synthesized properties across subject apps (RQ1).}
\label{tab:rq1_results}
\renewcommand{\arraystretch}{1.4} 
\begin{adjustbox}{max width=0.48\textwidth}
\begin{tabular}{c|cc|cc|cc|c|c}
\hline
\hline
\multirow{2}{*}[-2.5pt]{\textbf{App Name}}
& \multicolumn{2}{c|}{\textbf{\#Inferred Func}} 
& \multicolumn{2}{c|}{\textbf{\#Valid Func}} 
& \multicolumn{2}{c|}{\makecell[c]{\textbf{\#Correctly Executed}}} 
& \multirow{2}{*}[-2.5pt]{\textbf{\#Prop}}
& \multirow{2}{*}{\makecell[c]{\textbf{\#Valid} \\ \textbf{Prop}}} \\
\cline{2-7} 
& \textbf{D} & \textbf{P} & \textbf{D} & \textbf{P} & \textbf{D} & \textbf{P} & & \\
\hline

\rowcolor[HTML]{D9EAD3}
\omninotes     & 53 & 108 & 53 (100.0\%) & 102 (94.4\%) & 18 (34.0\%) & 80 (74.1\%)  & 81  & 80 (98.8\%) \\ 

\markor        & 48 & 117 & 36 (75.0\%)  & 116 (99.1\%) & 12 (25.0\%) & 85 (72.6\%)  & 88  & 83 (94.3\%) \\ 

\rowcolor[HTML]{D9EAD3}
\retromusic    & 48 & 113 & 42 (87.5\%)  & 104 (92.0\%) & 9 (18.8\%)  & 88 (77.9\%)  & 89  & 70 (78.7\%) \\ 

\amaze         & 37 & 84  & 24 (64.9\%)  & 83 (98.8\%)  & 8 (21.6\%)  & 64 (76.2\%)  & 57  & 52 (91.2\%) \\

\rowcolor[HTML]{D9EAD3}
\myexpense    & 54 & 106 & 54 (100.0\%) & 104 (98.1\%) & 21 (38.9\%) & 82 (77.4\%)  & 81  & 78 (96.3\%) \\

\antennapod    & 47 & 142 & 38 (80.9\%)  & 139 (97.9\%) & 28 (59.6\%) & 116 (81.7\%) & 120 & 111 (92.5\%) \\ 

\rowcolor[HTML]{D9EAD3}
\anki    & 50 & 116 & 28 (56.0\%)  & 107 (92.2\%) & 9 (18.0\%)  & 93 (80.2\%)  & 91  & 85 (93.4\%) \\

\outertune     & 41 & 76  & 41 (100.0\%) & 67 (88.2\%)  & 9 (22.0\%)  & 55 (72.4\%)  & 48  & 44 (91.7\%) \\ 

\rowcolor[HTML]{D9EAD3}
\newpipe       & 46 & 106 & 45 (97.8\%)  & 92 (86.8\%)  & 20 (43.5\%) & 81 (76.4\%)  & 79  & 72 (91.1\%) \\

\files & 47 & 110 & 44 (93.6\%)  & 103 (93.6\%) & 23 (48.9\%) & 82 (74.5\%)  & 86  & 82 (95.3\%) \\

\rowcolor[HTML]{D9EAD3}
\orgzly        & 49 & 124 & 48 (98.0\%)  & 115 (92.7\%) & 14 (28.6\%) & 88 (71.0\%)  & 96  & 91 (94.8\%) \\

\uhabits        & 55 & 80  & 38 (69.1\%)  & 78 (97.5\%)  & 16 (29.1\%) & 63 (78.8\%)  & 69  & 64 (92.8\%) \\

\hline
\rowcolor[HTML]{EFEFEF} 
\textbf{Total} & \textbf{575} & \textbf{1282} & \textbf{491 (85.4\%)} & \textbf{1210 (94.4\%)} & \textbf{187 (32.5\%)} & \textbf{977 (76.2\%)} & \textbf{985} & \textbf{912 (92.6\%)} \\
\hline
\hline
\end{tabular}
\end{adjustbox}
\end{table}

\subsection{Results of RQ1}

Table~\ref{tab:rq1_results} reports the results of functionality inference, execution, and property synthesis on all subject apps, where \textit{\#Inferred Func} denotes the number of functionalities inferred by tools,  \textit{\#Valid Func}, \textit{\#Correctly Executed} and \textit{\#Valid Prop} denote the numbers of functionalities and properties validated by human annotators, and \textit{\#Prop} denotes the number of generated properties. 
For functionality-related evaluation, we compare DroidAgent (D) and our approach (P). 
Overall, our approach consistently outperforms DroidAgent in both functionality inference and execution. 
Across the 12 apps, our approach infers 1,282 functionalities, of which 1,210 are judged valid, achieving 94.4\% functionality validity, compared with 575 inferred functionalities and 491 valid ones (85.4\%) for DroidAgent. 
It also correctly executes 977 functionalities, yielding 76.2\% execution correctness, substantially higher than DroidAgent's 187 correctly executed functionalities and 32.5\% execution correctness. 
On average, this corresponds to 101 valid functionalities and 81 correctly executed functionalities per app for our approach, compared with 41 and 16, respectively, for DroidAgent.

One likely reason for the performance gap is the difference in functionality inference. 
DroidAgent infers functionalities without explicitly grounding them to concrete GUI widgets, making some inferred results loosely related to the current interface context and thus more likely to be invalid. 
In contrast, our approach grounds functionality inference in the current GUI context, which helps produce more valid functionality hypotheses and makes subsequent execution more reliable. Also, we find DroidAgent tends to repeatedly execute failed events during functionality execution.

For property synthesis, our approach generates 985 property descriptions, among which 912 are judged valid, corresponding to 92.6\% property validity.
Moreover, property validity exceeds 90\% on most apps, indicating that the synthesized properties are generally well aligned with observed app behaviors.
\revised{In contrast, the naive LLM baseline, which is given only the app description, produces 240 properties across the 12 apps, \emph{none} of which is valid: without observing the app at runtime, the LLM hallucinates functionalities and widgets that do not exist in the app, so the resulting properties reference non-existent UI elements and are not executable. The LLM alone is thus insufficient without runtime behavioral evidence.}

\noindent\textbf{\revised{Failure analysis.}}
\revised{\emph{Functionality validity.} The MLLM occasionally proposes a functionality that does not exist in the app, likely because the semantic context for inference (\S~\ref{sec:hypothesis_generation}) is sometimes insufficient. An \emph{interact-then-infer} strategy, which first probes a widget and then infers the functionality from the observed effect, could help.
\emph{Execution correctness.} Failed executions either do not complete the functionality within the step budget, or are prematurely labeled as successful. The former could be mitigated by an activity transition graph derived via static analysis to guide step planning, the latter by an explicit completion check grounded in post-state evidence rather than the MLLM's own judgment.
\emph{Property validity.} The LLM occasionally hallucinates widgets absent from the GUI when synthesizing $P$, $I$, or $Q$, despite the grounding requirement (\S~\ref{sec:property-synthesis}). A post-synthesis check that verifies every referenced widget against the source trace could reject or regenerate such properties.}

\noindent\textbf{LLM usage cost.}
For \textit{behavioral evidence construction} and \textit{property synthesis}, each LLM call consumes 6,198 tokens / \$0.0145 on average. 
Overall, our approach uses 6,364k tokens / \$14.86 per app on average.
In comparison, DroidAgent uses 5,109k tokens / \$13.01 per app.

\begin{table}[t]
\centering
\small
\caption{Effectiveness of property refinement across subject apps (RQ2).}
\label{tab:rq2_results}
\renewcommand{\arraystretch}{1.3} 
\begin{adjustbox}{max width=0.45\textwidth}
\begin{tabular}{c|c|c|c|c|c}
\hline
\hline
\multirow{2}{*}[-2pt]{\textbf{App Name}} 
&\multirow{2}{*}[-2pt]{\textbf{\#Imprecise Prop}} 

& \multirow{2}{*}[-2pt]{\textbf{\#Successful Refinements}}

& \multicolumn{3}{c}{\textbf{Modified Component}} \\ 
\cline{4-6} 
 & & & \textbf{Pre} & \textbf{\quad I \quad} & \textbf{Post} \\
\hline

\rowcolor[HTML]{FCE5CD}
\omninotes     & 12 & 11 (91.7\%)  & 5 & 0 & 6 \\
\markor        & 15 & 12 (80.0\%)  & 2 & 1 & 9 \\

\rowcolor[HTML]{FCE5CD}
\retromusic    & 15 & 14 (93.3\%)  & 9 & 0 & 5 \\
\amaze         & 11 & 10 (90.9\%)  & 2 & 0 & 8 \\

\rowcolor[HTML]{FCE5CD}
\myexpense    & 8  & 8 (100.0\%)  & 6 & 0 & 2 \\
\antennapod    & 20 & 18 (90.0\%)  & 7 & 0 & 11 \\

\rowcolor[HTML]{FCE5CD}
\anki     & 9  & 9 (100.0\%)  & 7 & 0 & 2 \\
\outertune     & 3  & 3 (100.0\%)  & 3 & 0 & 0 \\

\rowcolor[HTML]{FCE5CD}
\newpipe      & 11 & 10 (90.9\%)  & 2 & 0 & 8 \\
\files & 6  & 6 (100.0\%)  & 2 & 0 & 4 \\

\rowcolor[HTML]{FCE5CD}
\orgzly        & 12 & 12 (100.0\%) & 5 & 0 & 7 \\
\uhabits        & 5  & 5 (100.0\%)  & 3 & 0 & 2 \\

\hline
\rowcolor[HTML]{EFEFEF} 
\textbf{Total} & \textbf{127} & \textbf{118 (92.9\%)} & \textbf{53} & \textbf{1} & \textbf{64} \\
\hline
\hline
\end{tabular}
\end{adjustbox}
\vspace{-10pt}
\end{table}

\subsection{Results of RQ2}

Table~\ref{tab:rq2_results} reports the effectiveness of property refinement across all subject apps, including how many generated properties produce false positives during property-based testing, how many are successfully refined, and which property components are modified.

Among all generated executable properties, 127 produce spurious violations during testing and are thus identified as imprecise properties that lead to false positives. 
This result shows that property imprecision is not uncommon in LLM-generated properties, and therefore refinement is necessary to improve their practical usability.
Overall, our refinement technique is highly effective. 
Across all apps, 127 properties are sent to refinement, and 118 of them are successfully refined, yielding an overall refinement rate of 92.9\%.
Moreover, the refinement performance is consistently strong across apps: six apps achieve a 100\% refinement rate, while the remaining apps still achieve rates at least 80\%.

We further analyze which property components are modified during refinement. 
Among the 118 successfully refined properties, 53 involve precondition modifications, 64 involve postcondition modifications, and only 1 involves an interaction modification. 
This suggests that most false positives can be resolved by refining when a property should be triggered or what outcome it should assert, rather than changing the core interaction sequence.
Among the 127 imprecise properties, 9 remain not refined.
We find that 5 of them are caused by incorrect diagnosis of spurious violations, while the others occur because the revised property can no longer remain consistent with the original functionality.
\revised{More specifically, in the 5 misdiagnosed cases the refinement attributes the false positive to the wrong component among $P$, $I$, and $Q$, and therefore revises a component that is not the actual cause; in the remaining 4 cases the revision drifts away from the original testing intent.
These two patterns suggest two directions for improvement: providing richer context (\eg, more detailed traces or intermediate GUI states) to help localize the true cause of a false positive, and adding a post-refinement intent-verification step (\eg, an LLM judge against the source trace) that rejects or regenerates revisions that drift from the original intent.}
Note that for \textit{property refinement}, each property requires 8,768 tokens / \$0.02 on average.

\subsection{Results of RQ3}

\begin{table}[!t]
\centering
\small
\caption{Statistics of the \findbugs new functional bugs found by the generated properties.}
\label{table:bug_found}
\renewcommand{\arraystretch}{1.2}
\newcommand{\bugrowcolor}{\cellcolor[HTML]{FFF2CC}} 
\begin{adjustbox}{max width=0.48\textwidth}
\begin{tabular}{c|c|l}
\hline
\hline
\textbf{App Name} & \textbf{ID} & \textbf{Violated Property} \\ \hline

\multirow{10}{*}{\omninotes} & \bugrowcolor 1  & \bugrowcolor Note info dialog should contain statistical data \\
                             & 2  & The date should appear on the selection page \\
                             & \bugrowcolor 3  & \bugrowcolor The category selection should be changeable \\
                             & 4  & The category page should be accessible from the drawer \\
                             & \bugrowcolor 5  & \bugrowcolor The captured photo should appear in the note content \\
                             & 6  & Returning from the sketch should display the note title \\
                             & \bugrowcolor 7  & \bugrowcolor The reminder icon should be displayed after setting a reminder \\
                             & 8  & The attachment should appear after selection \\
                             & \bugrowcolor 9  & \bugrowcolor The image can be opened in the note content \\
                             & 10 & The FAB should appear after return \\
\hline

\multirow{6}{*}{\retromusic} & \bugrowcolor 11 & \bugrowcolor The specific item should disappear from the list \\
                             & 12 & The lyrics should be displayed after saving \\
                             & \bugrowcolor 13 & \bugrowcolor The item should not exist in the list \\
                             & 14 & The image should be displayed after change \\
                             & \bugrowcolor 15 & \bugrowcolor The artist can be reset to default \\
                             & 16 & The metadata should remain consistent \\
\hline

\amaze                       & \bugrowcolor 17 & \bugrowcolor The item should be deleted successfully \\
\hline

\outertune                   & 18 & The song list should open from navigation menu \\
\hline

\multirow{3}{*}{\files}      & \bugrowcolor 19 & \bugrowcolor The file can be successfully created \\
                             & 20 & The item can be found \\
                             & \bugrowcolor 21 & \bugrowcolor Navigation to the sub-directory should succeed \\
\hline

\uhabits                     & 22 & The added number should keep consistent \\
\hline

\orgzly                      & \bugrowcolor 23 & \bugrowcolor The imported file should be present \\
\hline

\multirow{2}{*}{\markor}     & 24 & Star and ``Favourite'' checkbox should stay in sync \\
                             & \bugrowcolor 25 & \bugrowcolor The insert image should be displayed after preview \\
\hline
\hline
\end{tabular}
\end{adjustbox}
\vspace{-10pt}
\end{table}

Table~\ref{table:bug_found} summarizes the bug-finding results, including the app name, bug ID, and the brief description of the violated property. 
Overall, \tool uncovered \findbugs unique previously unknown functional bugs in the latest app versions. Currently, 5 of the reported bugs have been fixed by developers, while the remaining reports are waiting for responses.

\revised{The bugs span four categories: (i) \emph{functionality does not take effect} (12 bugs, \eg, a captured photo is not attached to the note, or a deleted item remains in the list), (ii) \emph{state inconsistencies} (6 bugs, \eg, the reminder icon is not shown after a reminder is set, or the lyrics of a song are not updated after saving), (iii) \emph{unexpected wrong behaviors} (4 bugs, \eg, the floating action button fails to reappear after returning to the note list, or an info dialog does not contain the expected statistical data), and (iv) \emph{navigation failures} (3 bugs, \eg, a page is unreachable from the drawer).
The bugs also cover diverse functionalities (\eg, note management, attachment insertion, content display), suggesting that the generated properties are not tied to a single type of expected behavior and can capture a broad range of behavioral constraints.}
These bugs arise when the actual app behavior deviates from the expected behavior encoded by the properties.
For example, Bug 5 in \textit{OmniNotes} is triggered by the property \textit{The captured photo should appear in the note content}. 
During testing, it generates a GUI event sequence that first opens audio recording and then checks this property. 
After executing the interaction sequence, the captured photo fails to appear, thus violating the property.
This bug is difficult to uncover through manual testing, as testers typically focus on the main interaction path and may not consider interleavings with other events.

\begin{table}[!t]
	\centering
	\small
	\caption{\revised{The \findbugs new bugs found by \tool \wrt prior testing tools: how many fall within each tool's detection scope, and how many it rediscovers in practice. Totals are deduplicated across tools.}}
	\label{table:existingtool}
	\renewcommand{\arraystretch}{1.0} 

	\setlength{\tabcolsep}{12pt}

	\begin{adjustbox}{max width=0.48\textwidth}
		\begin{tabular}{c|c|c} 
			\hline
			\hline
			\textbf{Tool} & \textbf{\#In Scope} & \textbf{\#Rediscovered} \\ \hline

			\rowcolor[HTML]{CFE2F3}
			\textsc{Genie}       & 1 (4.0\%)   & 0 (0.0\%) \\
			\textsc{ODIN}        & 2 (8.0\%)   & 0 (0.0\%) \\

			\rowcolor[HTML]{CFE2F3}
			\textsc{PBFDroid}    & 5 (20.0\%)  & 2 (8.0\%) \\
			\textsc{VisionDroid} & -           & 1 (4.0\%) \\

			\hline
			\rowcolor[HTML]{EFEFEF}
			\textbf{Total}     & \textbf{7 (28.0\%)} & \textbf{3 (12.0\%)} \\ \hline
			\hline
		\end{tabular}
	\end{adjustbox}
\end{table}

\subsection{Results of RQ4}
Table~\ref{table:existingtool} reports how the \findbugs new bugs uncovered by \tool relate to the four prior techniques: only 7 (28\%) are within their scope, and only 3 (12\%) are actually found in practice.
Genie, Odin, and PBFDroid rely on predefined automated oracles or manually specified DMF properties, and can therefore only find the bug categories emphasized in their original designs, whereas \tool targets more broadly.
VisionDroid, in contrast, is a more general LLM-based approach, but its exploration validates a functionality by following one plausible interaction path at a time, whereas most bugs uncovered by \tool are exposed only when a property is checked under specific event sequences.

\revised{We further compare the overall bug-finding results. On the same 12 apps, the four prior techniques together uncover 13 previously unknown functional bugs (Genie 3, Odin 3, PBFDroid 6, and VisionDroid 4; the per-tool counts sum to 16 because three bugs are found by more than one tool), fewer than the \findbugs bugs uncovered by \tool.
Both counts follow the same criterion, \ie, every bug is manually confirmed to be a real, previously unknown functional bug; as reported in RQ3, 5 of those found by \tool have already been fixed by developers.
The 3 rediscovered bugs above are exactly the overlap between the two bug sets: \tool finds 22 bugs that the prior techniques miss, while they find 10 that \tool misses.
The two are thus complementary rather than substitutable, as both directions stem from the oracles each side can check.
\tool misses those 10 bugs mainly because its generated properties do not cover the oracles these techniques rely on, \eg, the differential oracles of Genie and Odin; it also generates only one property per functionality, and generating more would cover more expected behaviors.}

\revised{The two bug sets also differ in type. The bugs found by each prior technique are concentrated in the category its oracle targets: all bugs found by PBFDroid are data manipulation errors; Genie and Odin mainly expose unexpected wrong behaviors; and VisionDroid mainly exposes display errors. In contrast, the bugs found by \tool span four categories reported in RQ3.
\tool therefore achieves broader coverage without manual effort, but does not subsume a specialized oracle within its own niche; combining the two is a promising direction for future work.}

\section{Discussion}

\noindent{\textbf{\revised{Impact of pre-existing app faults on generated properties.}}}
\revised{A pre-existing app fault could in principle corrupt the properties \tool infers or refines.
In the \emph{generation} phase, the risk is limited because \tool synthesizes properties only from successfully executed traces, so a fault that blocks a functionality yields no property, and it abstracts the observed behavior into a general property rather than copying a concrete trace.
In the \emph{refinement} phase, a correct property could be weakened only if a fault-induced violation is misjudged as spurious; violations attributed to app faults are reported as bugs instead, and refinement yields only a \emph{plausibly correct} property whose final correctness is confirmed by humans. }

\smallskip
\noindent{\textbf{Threats to validity.}}
First, our evaluation involves manual inspection to assess the correctness of inferred functionalities and generated properties, which may introduce subjectivity and potential bias. To mitigate this threat, each case is independently labeled by two experienced graduate students following consistent evaluation criteria; disagreements are further discussed until agreement is reached.
Second, the apps used in our evaluation may not fully represent the diversity of real-world mobile apps. To mitigate this threat, most of the apps are selected from prior relevant studies, and we further include four additional apps to improve diversity. In the future, we plan to evaluate \tool on a larger and broader set of apps.
\revised{Third, the subject apps are open-source and may appear in GPT-5.2's training data, which raises the concern that \tool relies on the model's prior knowledge of these apps rather than genuinely inferring the bugs from the collected runtime evidence. To mitigate this threat, we checked each of the \findbugs bugs against the corresponding app's issue tracker and confirmed that none had been previously reported. Moreover, six of the 12 app builds we tested were released after GPT-5.2's training cutoff, and \tool still finds bugs in them.}

\section{Related Work}


\noindent{\textbf{Automated mobile app GUI testing.}}
Automated testing for mobile apps has been extensively studied.
Choudhary \etal~\cite{choudhary2015automated} conducted a systematic comparison of Android input-generation tools and highlighted both the promise and limitations of automated mobile testing.
Prior work has proposed a variety of techniques to automatically explore app GUIs and generate event sequences for detecting crash bugs~\cite{su_stoat_2017,dong2020time,gu2019practical,wang2020combodroid,machiry2013dynodroid,mao2016sapienz,wang2025llmdroid,liu2024make}.
For example, Sapienz~\cite{mao2016sapienz} uses multi-objective search to generate event sequences for improving coverage and exposing crashes.
To find non-crash functional bugs, most of prior work~\cite{genie,setdroid,setdroid_tse,dld,adamsen2015systematic,zaeem2014automated,xiong2023empirical,odin} designs \emph{automated oracles} to overcome the oracle problem. However, these work are limited to specific types of functional bugs (\eg, data losses~\cite{zaeem2014automated,adamsen2015systematic,dld}). Some work leverages LLM to analyze GUI pages during exploration to find data inconsistency bugs~\cite{hu2024autoconsis}, functional bugs~\cite{liu2025seeing}, or inconsistencies between app design and implementation~\cite{liu2025guipilot}.

Property-based testing (PBT) is a powerful testing methodology and has been adopted into many different software systems to find logic bugs~\cite{arts2006testing,2022quickstrom,hughes2016experiences,hughes2016mysteries,santos2018property,karlsson2020quickrest,padhye2019jqf,hypothesis,claessen2000quickcheck}.
Recent work has begun to bring property-based testing to mobile apps. 
Specifically, PBFDroid~\cite{pbfdroid}, PDTDroid~\cite{sun2024property}, and Kea~\cite{kea,liang2026kea2} have demonstrated that PBT can be effectively applied to GUI-driven mobile apps and can find non-crashing functional bugs that are difficult for traditional automated GUI testing tools to detect. However, these work still assumes that meaningful properties are manually written by developers or testers. Our work complements these approaches by automating the executable property generation.

\noindent{\textbf{Automated test \revised{and specification} generation.}}
Traditional automated test generation techniques, such as fuzzing~\cite{afl}, symbolic execution~\cite{godefroid2005dart,sen2005cute,tillmann2014transferring}, and search-based testing~\cite{fraser2011evosuite,pacheco2007randoop}, mainly aim to improve code coverage and expose crashes, but they struggle to produce effective assertions for functional correctness~\cite{panichella2020revisiting,shamshiri2015automated}.
\revised{To alleviate this oracle problem, one line of work mines likely specifications from program executions, learning value-based invariants (\eg, Daikon~\cite{ernst1999daikon}), temporal API rules (\eg, Perracotta~\cite{yang2006perracotta} and Texada~\cite{lemieux2015general}), or behavioral models and typestates (\eg, TzuYu~\cite{xiao2013tzuyu} and Synoptic~\cite{beschastnikh2011synoptic}).}
\revised{More recently, learning-based and LLM-based approaches generate unit tests and assertions directly from source code~\cite{tufano2020unit,lahiri2022interactive,yuan2024evaluating,yang2024evaluation,wang2024software,fan2023large,gao2025current,dozono2024large,jiang2024towards,deng2023large,schafer2023empirical,chen2024chatunitest}, and LLMs have been used to generate specifications or properties in specific domains, \eg, postconditions for functions from natural-language comments~\cite{endres2024can}, property-based tests from docstrings for Python libraries~\cite{vikram2023can}, and properties for smart contracts from a curated property library~\cite{liu2024propertygpt}.}

\revised{Like these works, \tool automatically infers behavioral specifications, but differs from all of them in both setting and methodology. All the above start from a given unit under analysis together with a structured source of intent, either human-written specifications such as docstrings, comments, or a curated property library, or white-box execution data such as method entry/exit values and API call traces. \tool has neither: it must first \emph{discover} which functionalities to specify by exploring a black-box app, and infers their intended behavior from runtime GUI states, screenshots, and event sequences alone. It further refines the synthesized properties against testing feedback, which these works do not.}

Beyond test and specification generation, recent work proposes agents that automatically perform tasks on mobile apps~\cite{wen2024autodroid,wen2023droidbot,zhang2025appagent,wang2024mobile,qin2025ui,yoon2024intent,ran2024guardian}. These agents focus on executing user-provided tasks, whereas \tool centers on automatically exploring app functionalities and generating executable properties for testing.
\section{Conclusion}
In this paper, we present \tool, an automated approach for constructing properties of mobile apps. 
By exploring app functionalities and deriving properties from behavioral evidence, \tool reduces the need for manual property specification. 
We further propose a feedback-driven refinement technique to refine imprecise properties exposed during testing. 
Experiments on real-world Android apps show that \tool can effectively generate valid executable properties.
These results demonstrate the practical value of automated property construction for mobile app property-based testing.
\section{Data Availability}
We have made all the artifacts publicly available~\cite{propgen_artifact}.
\begin{acks}
We thank the anonymous ASE reviewers for their valuable feedback. This research is supported by the National Research Foundation, Singapore and CyberSG R\&D Programme Office under its Translation and Innovation Grant (CRPO-GC4-SMU-002). Any opinions, findings and conclusions or recommendations expressed in this material are those of the author(s) and do not reflect the views of the National Research Foundation and CyberSG R\&D Programme Office. This work was also partially supported by NSFC Project No. 62572192 and 62072178.
\end{acks}
\newpage

\bibliographystyle{ACM-Reference-Format}
\bibliography{ref.bib}

\end{document}